%% file: main.tex
\documentclass[11pt]{article}

\usepackage[preprint]{acl}
\usepackage{times}
\usepackage{latexsym}
\usepackage[T1]{fontenc}
\usepackage[utf8]{inputenc}
\usepackage{microtype}
\usepackage{booktabs}
\usepackage{array}
\usepackage{tabularx}
\usepackage{graphicx}
\usepackage{stfloats}
\usepackage{placeins}
\usepackage{xcolor}
\usepackage{amsmath}
\usepackage{amssymb}
\usepackage[most]{tcolorbox}
\newcolumntype{Y}{>{\raggedright\arraybackslash}X}

\definecolor{promptbluebg}{HTML}{F3F7FF}
\definecolor{promptblueframe}{HTML}{4C6FB3}
\definecolor{promptredbg}{HTML}{FFF4F4}
\definecolor{promptredframe}{HTML}{B94A48}
\definecolor{promptgreenbg}{HTML}{F4FFF7}
\definecolor{promptgreenframe}{HTML}{4A8A5B}
\definecolor{takeawaybg}{HTML}{F0FBFA}
\definecolor{takeawayrule}{HTML}{168C83}
\definecolor{takeawaytext}{HTML}{123A38}

\newtcolorbox{promptbox}[2][]{
  enhanced,
  breakable,
  colback=promptbluebg,
  colframe=promptblueframe,
  coltitle=black,
  fonttitle=\bfseries,
  title={#2},
  boxrule=0.9pt,
  arc=1.5mm,
  left=1.2mm,
  right=1.2mm,
  top=1mm,
  bottom=1mm,
  #1
}

\newtcolorbox{keyfindingbox}{
  enhanced,
  breakable,
  colback=takeawaybg,
  colframe=takeawaybg,
  boxrule=0pt,
  borderline west={1.1mm}{0pt}{takeawayrule},
  arc=0.8mm,
  left=2.0mm,
  right=1.2mm,
  top=0.8mm,
  bottom=0.8mm,
  fontupper=\small\color{takeawaytext}
}
\newcommand{\keyfinding}[1]{%
  \begin{keyfindingbox}
  \textbf{\textsc{Takeaway.}} #1
  \end{keyfindingbox}
}
\makeatletter
\providecommand{\phantomsection}{}
\newcommand{\appsubsection}[3]{%
  \subsection*{#1 #2}%
  \phantomsection
  \def\@currentlabel{#1}%
  \label{#3}%
}
\makeatother
\title{Louder, Longer, Livelier: Acoustic Shortcuts and Underspecified Rationales in Speech LLM Judges}
\author{
\makebox[\dimexpr\textwidth-2\tabcolsep\relax][c]{%
Mingyue Huo$^{1}$\textsuperscript{*}\quad
Shivam Mehta$^{2}$\quad Bhavin Jawade$^{2}$\quad Yinghong Lan$^{2}$\quad Haoqi Li$^{2}$}\\
\normalfont\makebox[\dimexpr\textwidth-2\tabcolsep\relax][c]{%
$^{1}$University of Illinois Urbana-Champaign \qquad $^{2}$Netflix}\\[-1pt]
\normalfont\small\makebox[\dimexpr\textwidth-2\tabcolsep\relax][c]{%
\texttt{mhuo5@illinois.edu} \qquad \texttt{haoqil@netflix.com}}
}

\begin{document}
\maketitle
\begingroup
\renewcommand{\thefootnote}{*}
\footnotetext[1]{Work done during internship at Netflix.}
\endgroup

\begin{figure*}[!b]
  \centering
  \includegraphics[width=\textwidth,trim={40bp 220bp 40bp 168bp},clip]{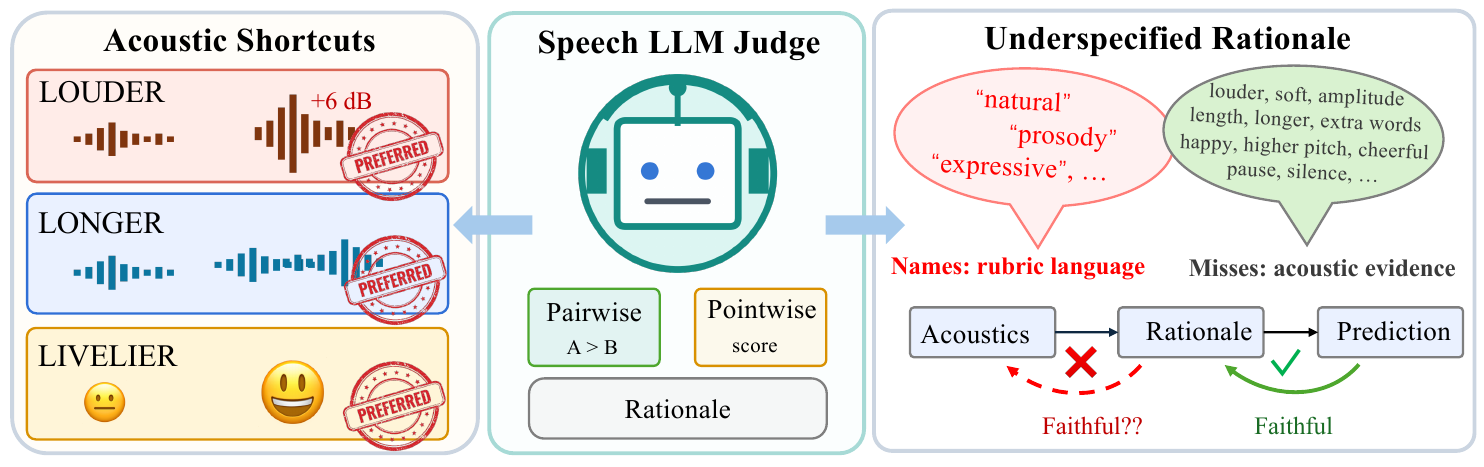}
  \vspace{-0.8em}
  \caption{Speech LLM judges may turn salient acoustic cues such as louder, longer, or livelier speech into quality preferences, while their rationales often use broad rubric terms rather than naming the cue that changed the judgment.}
  \label{fig:teaser}
\end{figure*}

\begin{abstract}
LLM-as-a-judge is widely used for evaluating text, but extending this paradigm
to speech requires models to interpret acoustic as well as linguistic evidence.
This introduces a modality-specific risk: a speech judge may treat a
perceptually salient cue as evidence of quality even when that cue is irrelevant
to the target criterion or receives more weight than human listeners give it.
We call this behavior an \textit{acoustic shortcut}.
To study it, we audit six speech LLM judges using controlled manipulations of
intensity, content richness, and emotional delivery. We evaluate both pointwise
scoring and pairwise comparison, using human preference calibration to interpret
the results.
The judges consistently reward louder audio, prefer content-rich speech more
strongly than human listeners do, and map emotional delivery into quality
preferences. These effects are most visible in pairwise comparison, while
pointwise scores often obscure them.
More concerningly, the accompanying rationales rarely identify the acoustic cue
that changes a judgment and instead repeatedly rely on a limited vocabulary,
leaving them acoustically underspecified.
Together, these findings show that reliable speech judges must both resist
acoustic shortcuts and ground their rationales in the acoustic evidence behind
their decisions.
To support reproducibility and future audits, we also release
\href{https://huggingface.co/datasets/mingyue66/SpeechJudgeAudit/}{\textsc{SpeechJudgeAudit}},
the controlled stimuli and evaluation tools used in this study.
\end{abstract}

\input{sections/01_intro}
\input{sections/02_related_work}
\input{sections/03_setup}
\input{sections/04_results}

\input{sections/05_rationales}
\input{sections/06_implications}
\input{sections/09_conclusion}
\input{sections/08_limitations}

\bibliography{references}

\appendix
\input{sections/appendix}

\end{document}

%% file: sections/01_intro.tex
\section{Introduction}
\label{sec:intro}
Speech quality is multidimensional and requires fine-grained acoustic
perception. Listeners may attend to clarity, intelligibility,
pleasantness, listening effort, naturalness, and other attributes depending on
the task and context~\citep{loizou2011speechquality}. Human listening tests such as Mean
Opinion Score (MOS) and pairwise studies remain the most direct way
to measure these judgments, but they are slow, expensive, and difficult to run
at every stage of speech technology development~\citep{wester2015we,kirkland2023stuck}.
Neural MOS predictors support faster iteration, but they are usually trained to
reproduce a particular scalar rating, provide little information about which
acoustic evidence shaped that score, and may not generalize to new listening-test
contexts~\citep{cooper2022generalization}.

Speech LLM judges offer another way to scale evaluation. They range from
specialized large audio language models (LALMs) fine-tuned for speech-quality
evaluation to proprietary, general-purpose LALMs evaluated with zero-shot
prompts. They can listen to one or two audio clips and return a score,
preference, or descriptive rationale. Specialized judges also differ in scope:
SpeechJudge targets
naturalness~\citep{speechjudge2025}, while SQ-LLM~\citep{sqllm2025} and
UniSRM~\citep{unisrm2026} assess multiple dimensions of overall quality,
including intelligibility, prosody, and listening effort. Their outputs can
support data selection and inference-time evaluation, or serve as reward
signals for training speech models~\citep{speechjudge2025,wavreward2025}.
Consequently, biases in these judgments can affect both leaderboard results and models
trained with judge feedback.

This raises a fundamental question: do speech LLM judges evaluate the intended
quality reliably, or do they let salient acoustic cues become shortcuts?
Text-based LLM judges are known to favor candidates based on their position,
length, or model family~\citep{zheng2023judging,wang2023fair,
saito2023verbosity,panickssery2024selfrecognition}. Speech LLM
judges may inherit these risks, while the audio modality introduces a distinct,
modality-specific risk: speech varies in pitch, loudness, pauses, speaking rate,
emotional delivery, and more, and a judge must determine which of these cues are
relevant to the quality being evaluated.

To study this risk, we examine whether judges use acoustic cues in
ways justified by the intended quality criterion. \textbf{We call this behavior
an \emph{acoustic shortcut} when an easy-to-detect acoustic cue shapes a quality
judgment despite being irrelevant to the target criterion or receiving much
more weight than human listeners give it.}
We audit six speech LLM judges using both human recordings and generated speech.
Although one audited model is a LALM-based scalar reward model rather than a
generative rationale-producing judge, we include it as a speech-LLM-based
evaluator.
Our controlled comparisons isolate one factor at a time,
allowing us to test whether each manipulation systematically affects
final judgments and, when it does, whether the accompanying rationales identify
the relevant acoustic change. We use human preference calibration when a factor
may reasonably affect perceived quality.
Our contributions and findings are:
\begin{itemize}
  \item \textbf{A controlled acoustic audit of speech LLM judges.}
  We audit six judges using controlled manipulations of intensity, content
  richness, and emotional delivery, with human calibration when a factor may
  affect perceived quality. We test whether controlled acoustic changes affect
  pointwise and pairwise judgments and whether rationales name the relevant
  acoustic changes. We release the
  resulting stimuli and evaluation tools as \textsc{SpeechJudgeAudit}, allowing
  the same tests to be applied to other speech judges.
  Section~\ref{sec:setup} describes the methodology.
  
  \item \textbf{Judges turn salient acoustic cues into quality preferences.} We find that judges reward
  louder audio even when only signal level changes, prefer content-rich speech
  more strongly than human listeners do, and map emotional delivery into quality
  preferences. These effects are clearest in pairwise comparison, while
  pointwise scores often make them less visible. Section~\ref{sec:results}
  presents the results. Such judges can reward systems for
  changing easy-to-detect acoustic properties instead of improving the intended
  quality.

  \item \textbf{Rationales are acoustically underspecified.} We find that rationales repeatedly draw on a limited vocabulary
  across acoustically different inputs. Even when a manipulation changes the
  judgment, the rationales rarely name the manipulated acoustic cue, leaving
  them acoustically underspecified. Detailed results are presented in Section~\ref{sec:rationales}.
  This limits their value as debugging signals for speech evaluation and model development.

  \item \textbf{Implications for using and training speech judges.} The audit also
  supports practical guidance for speech evaluation: control salient acoustic
  properties before judging, compare pointwise and pairwise behavior when
  possible, and train future judges with rationale data grounded in localized
  acoustic evidence rather than only broad rubric language. Section~\ref{sec:implications}
  discusses these implications.
\end{itemize}

%% file: sections/02_related_work.tex
\section{Related Work}
\label{sec:related}

\subsection{Speech LLM Judges}
\label{sec:rel-judges}

Recent work has explored large audio language models (LALMs) as scalable
evaluators of speech. One line of work uses general-purpose LALMs without
task-specific fine-tuning. AudioJudge evaluates zero-shot speech assessment with
general-purpose audio models and documents reliability issues such as position
and transcript-length bias~\citep{audiojudge2025}. TRACE instead converts
acoustic measurements into a structured written description before asking a text
LLM to evaluate the speech~\citep{trace2026}.

Another line of work develops specialized judges by fine-tuning existing LALMs,
such as Qwen2.5-Omni~\citep{qwen25omni2025}, for speech-quality evaluation. SpeechJudge focuses on
naturalness and uses human preference data to train both a scalar
Bradley--Terry model and a pairwise judge that also produces
rationales~\citep{bradley1952rank,speechjudge2025}. The Bradley--Terry
baseline, SpeechJudge-BTRM, adds a linear reward head to Qwen2.5-Omni-7B and
outputs a single scalar reward; unlike SpeechJudge-GRM, it does not generate
rationales. SQ-LLM and UniSRM also build on the same backbone but target broader
speech-quality and reward-modeling settings~\citep{sqllm2025,unisrm2026}.
Related work further broadens speech evaluation to low-level quality
descriptions, spoken interaction, and instruction-driven
evaluation~\citep{qualispeech2025,wavreward2025,gsrm2026,jastin2026}.

These studies show that speech LLM judges can return scores,
preferences, and natural-language rationales. However, prior
work has not systematically tested whether speech judges treat controlled
acoustic cues as invalid or overweighted evidence of quality. Our work addresses
this question through controlled acoustic manipulations across both
general-purpose and specialized judges.

\subsection{From LLM-Judge Bias to Acoustic Shortcuts}
\label{sec:rel-text}

Research on text-based LLM judges provides a starting point for understanding
this problem. These judges can favor candidates based on position, length,
verbosity, or model identity rather than the target
quality~\citep{wang2023fair,zheng2023judging,singhal2023long,saito2023verbosity,
dubois2024length,panickssery2024selfrecognition,saito2024selfpreference}. When
judge outputs are used as training rewards, such biases may lead to reward
hacking~\citep{gao2022scaling}. Together, these are instances of shortcut
learning, in which an easily observed property shapes a judgment more than the
quality being evaluated~\citep{geirhos2020shortcut}.

The audio modality broadens the set of possible shortcuts because speech carries
linguistic content alongside loudness, pitch, duration, silence, speaking rate,
emotional delivery, and other acoustic properties~\citep{loizou2011speechquality}.
Controlled studies show that LALMs can favor easier linguistic or contextual
cues over acoustic evidence in emotion, paralinguistic, and speaker-consistency
tasks~\citep{listen2025,voxparadox2026,speakersleuth2026}. Although these
studies do not directly evaluate speech judges, they motivate auditing how
specific acoustic cues shape quality judgments.

\subsection{Rationale Faithfulness and Acoustic Evidence}
\label{sec:rel-listen}

Acoustic shortcuts concern what judges decide; rationales raise
a second question: whether those decisions are grounded in the audio. To make
judgments transparent, rationales should connect decisions to the relevant
acoustic evidence.

A growing body of work trains LALMs to produce multi-step rationales for audio
tasks, especially audio question answering~\citep{audiocot2025,
audioreasoner2025,afsoundcot2025,audsemthinker2025,thinkingwithsound2025,
r1aqa2025,audiodeepthinker2026}. These studies show that LALMs can produce
structured reasoning and that such reasoning can improve task performance.

For general audio reasoning,
\citet{lalmfaithful2025} intervene on model rationales and find that final
answers often depend on them. This tests \textit{rationale faithfulness}: whether
the answer follows the stated reasoning, regardless of whether that reasoning
accurately describes the audio. SpeechJudge reports similar internal
consistency in speech judging, with final preferences usually matching the
reasons stated in the rationale~\citep{speechjudge2025}.

These studies mainly test one link: whether the final prediction is
faithful to the stated rationale. Our audit asks about the other link: when a
controlled acoustic change affects a speech-judge decision, does the rationale
connect that decision to the acoustic evidence in the input? A rationale
can be consistent with the final preference while still failing to name the cue
that changed the judgment.

%% file: sections/03_setup.tex
\section{Methodology}
\label{sec:setup}

We conducted a controlled audit of six speech LLM judges. In each experiment,
we varied a targeted property of the speech while holding potential confounds
fixed, then examined (1) whether the change systematically affected final
judgments and (2) when it did, whether the accompanying rationales named
the relevant acoustic change.

\subsection{Judges and Audit Scope}
\label{sec:judges}

We selected six recent speech LLM judges, covering specialized models with
publicly available weights and general-purpose models evaluated
with zero-shot prompts. Their target criteria were naturalness or
overall speech quality, and their supported outputs included pointwise scores,
pairwise preferences, and rationales (Table~\ref{tab:judges}).

\begin{table*}[t]
\centering
\fontsize{8.5pt}{9.5pt}\selectfont
\setlength{\tabcolsep}{3.4pt}
\begin{tabular}{llllccc}
\toprule
\textbf{Judge} & \textbf{Backbone} & \textbf{Quality criterion} &
\textbf{Pairwise} & \textbf{Pointwise} & \textbf{Scale} &
\textbf{Rationale} \\
\midrule
SpeechJudge-BTRM~\citep{speechjudge2025}
  & Qwen2.5-Omni & Naturalness & $\times$
  & \checkmark & $(-\infty,\infty)$ & $\times$ \\
SpeechJudge-GRM~\citep{speechjudge2025}
  & Qwen2.5-Omni & Naturalness & \checkmark
  & $\times$ & 1--10 & \checkmark \\
UniSRM~\citep{unisrm2026}
  & Qwen2.5-Omni & Overall quality & \checkmark
  & \checkmark & 1--5 & \checkmark \\
SQ-LLM~\citep{sqllm2025}
  & Qwen2.5-Omni & Overall quality & \checkmark
  & \checkmark & 1--5 & \checkmark \\
Gemini-2.5-pro~\citep{gemini25report}
  & Gemini & Overall quality & \checkmark
  & \checkmark & 1--10 & \checkmark \\
Gemini-3.1-pro~\citep{gemini31modelcard}
  & Gemini & Overall quality & \checkmark
  & \checkmark & 1--10 & \checkmark \\
\bottomrule
\end{tabular}
\caption{Audited speech LLM judges. The Gemini models are general-purpose LALMs
evaluated with zero-shot prompts; the other four judges are specialized models
fine-tuned for speech-quality evaluation. SpeechJudge-BTRM is a scalar reward
model rather than a generative judge, but we include it as a LALM-based speech
evaluator. The rationale column indicates whether the judge returns an
explanation with its judgment.}
\label{tab:judges}
\end{table*}

The emotional-delivery experiment used human recordings. The remaining
experiments used synthesized stimuli from four text-to-speech (TTS) systems spanning
autoregressive and non-autoregressive architectures and both codec-token and
mel-spectrogram outputs. We used this diversity to avoid relying on a single
synthesis architecture. System details are provided in Appendix~\ref{app:tts}.

\subsection{Controlled Acoustic Manipulations}
\label{sec:manipulations}
We audit three controlled acoustic factors: intensity adjustment, content
richness, and emotional delivery. Complete stimulus construction details are
provided in Appendix~\ref{app:pools}.

\paragraph{Intensity.}
We selected 80 sentences from LibriTTS test-clean~\citep{libritts2019},
with 20 synthesized by each TTS system. We normalized each utterance to
$-30$ LUFS to form the fixed reference. For each shift
$g\in\{-6,-3,+3,+6\}$ dB, we scaled the reference signal by $10^{g/20}$,
changing its RMS level while holding all other signal properties fixed.

\paragraph{Content richness.}
We used 149 concise--richer pairs from ASSET~\citep{asset2020}. Each pair
expressed the same intended meaning and was synthesized with the same TTS system,
speaker, and reference voice. Appendix~\ref{app:pool-asset} gives a
representative concise--rich pair.

\paragraph{Duration controls.}
We used two controls to separate content richness from duration. For each ASSET
pair, silence padding matched the richer clip's duration without adding speech,
while repetition duplicated the concise speech without adding new content.
Table~\ref{tab:richness-decomp} reports the resulting comparison.

\paragraph{Emotional delivery.}
We used 800 recordings from
ESD~\citep{esd2022} where 10 speakers each recorded the same 20 sentences with Neutral,
Happy, Sad, and Angry delivery. Within each matched set, speaker and semantic
content were held constant, with intended emotional delivery as the experimental
factor. Before evaluation, we normalized all recordings to $-23$ LUFS and
trimmed leading and trailing silence.

\subsection{Evaluation and Analysis}
\label{sec:evaluation-analysis}

\paragraph{Judgment analysis.}
For pointwise evaluation, each clip was scored independently. Because judges
used different scales, we compared scores only within the same judge. Pairwise
evaluation selected one clip or returned a tie. We tested both A/B and B/A orders
to control position effects documented in text and audio
judges~\citep{wang2023fair,audiojudge2025}.
We encoded a manipulated-clip win as 1, a reference-clip win as 0, and a tie as
0.5, averaging the two orders within each item before aggregation. Confidence
intervals used item-level bootstrap samples. For pointwise emotional-delivery
scores, we fit within-judge mixed-effects models with emotion as a fixed effect
and random intercepts for speaker and sentence. Prompts, statistical tests, and
per-judge results are provided in Appendices~\ref{app:reproducibility}
and~\ref{app:stats}--\ref{app:mode-divergence}.

\paragraph{Human preference calibration.}
Fifteen native or near-native English listeners were assigned 30 pairs each,
yielding 449 valid judgments over 60 unique pairs: 12 intensity, 18
content-richness, 15 Sad--Neutral, and 15 Happy--Neutral. Presentation order was
randomized, and listeners selected the first clip, the second clip, or a tie.
Their responses provide a calibration
baseline: for content richness and emotional delivery, they help distinguish
judge-specific overweighting from preferences shared with listeners; for
intensity adjustment, they confirm that loudness is perceptually salient, even
though it is not valid evidence of better speech quality in our controlled
comparison.
Full instructions and response statistics are
provided in Appendix~\ref{app:human-details}.

\paragraph{Rationale analysis.}
The five rationale-producing judges returned either free-form explanations or
structured reasoning fields. Rationale analysis asks whether these explanations
make changed judgments transparent. We analyze two properties:
whether rationales adapt their vocabulary when acoustic inputs change, and
whether they explicitly name the manipulated acoustic cue when that cue affects
the judgment. Using prespecified lexicons for loudness, silence, repetition,
content richness, and emotion, we measure how often rationales name each
manipulation. We also compare frequent words across conditions to identify
recurring vocabulary. Judge--condition coverage and lexical summaries are
provided in Appendix~\ref{app:cot-lexicons}. This analysis concerns what the
rationales explicitly say, not the model's unobserved internal reasoning.

%% file: sections/04_results.tex
\section{Acoustic Shortcuts in Judge Decisions}
\label{sec:results}

For each acoustic manipulation, we ask whether it changes speech LLM judge
decisions, then compare judge behavior with human preference calibration when
the cue may plausibly affect perceived quality. This lets us distinguish cues
that are invalid evidence for the target quality from cues that judges
overweight relative to human listeners.

\subsection{Judges Reward Louder Audio}
\label{sec:res-loud}

\begin{figure}[t]
  \centering
  \includegraphics[width=0.8\columnwidth]{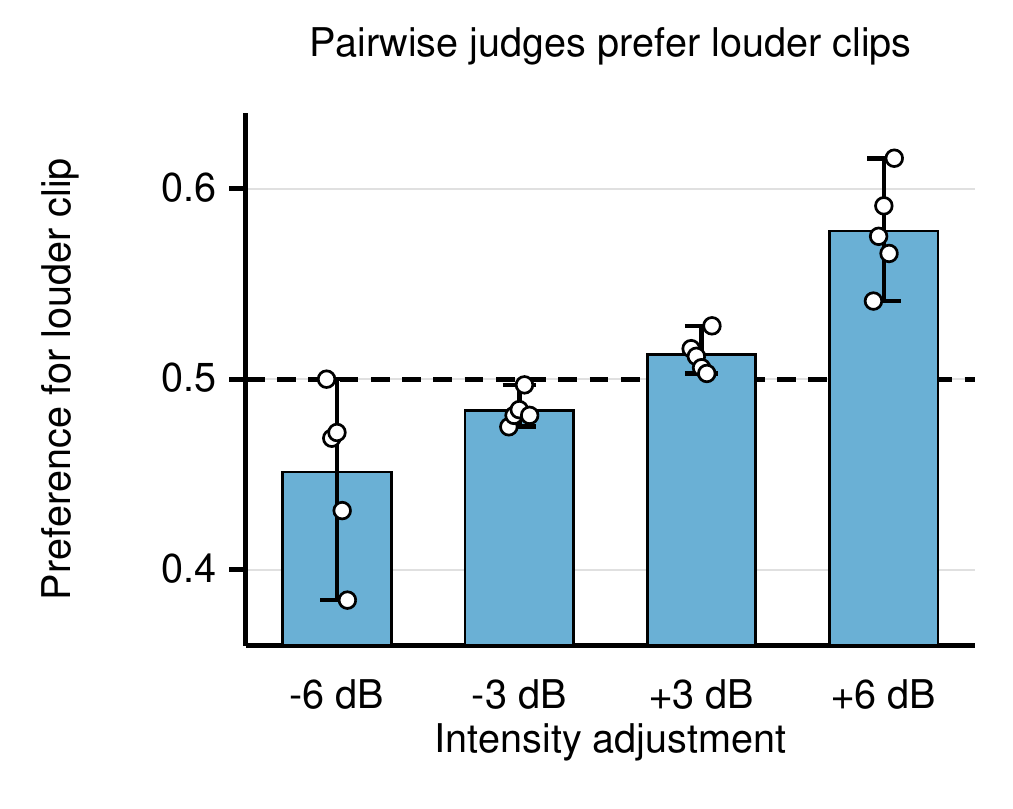}
  \caption{Mean pairwise preference for the louder clip after intensity adjustment. Points show individual judges; full per-judge results are in Appendix~\ref{app:loudness-full}.}
  \label{fig:loudness-pref}
\end{figure}

The intensity manipulation changes only the RMS level of a fixed reference
signal (Section~\ref{sec:manipulations}). Pairwise comparison reveals a
consistent preference for louder audio. At $+6$ dB, all five pairwise judges
favor the louder clip, with mean preference 0.578 and individual rates
from 0.541 to 0.616 (Figure~\ref{fig:loudness-pref}).

Pointwise scoring gives a less consistent picture. UniSRM assigns a small but
significant increase at $+6$ dB, whereas SQ-LLM returns identical pointwise
scores despite preferring the louder clip pairwise. The Gemini pointwise
differences are not significant, and BTRM penalizes quieter audio without
rewarding the $+6$ dB condition. Thus the directional loudness preference is
much clearer in pairwise comparison.

Human listeners also favor the $+6$ dB clip (0.612; Appendix~\ref{app:human-details}), confirming that the
change is perceptually salient. This does not make loudness valid evidence of
better speech quality: the intensity adjustment changes neither the linguistic
content nor the underlying speech. The judges can therefore reward louder audio
even when quality has not improved.
If such a judge is used for model selection or as a reward signal, it could
favor systems that increase output loudness without improving the intended
quality.

\keyfinding{All five pairwise judges favor the $+6$ dB speech: loudness is a clear acoustic shortcut.}

\subsection{Judges Overweight Content Richness}
\label{sec:res-richness}

\begin{table*}[t]
\centering
\small
\begin{tabularx}{\textwidth}{Y Y l}
\toprule
\textbf{Comparison} & \textbf{What changes} &
\textbf{Mean preference} \\
\midrule
Silence padding vs. concise & Duration only & 0.41 \\
Repetition vs. concise & Duplicated speech & 0.46 \\
Richer vs. concise & Content and varied speech material & 0.71 \\
Richer vs. duration-matched padding & Content richness at matched duration & 0.70 \\
\bottomrule
\end{tabularx}
\caption{Content-richness controls averaged over the five pairwise judges.
Values are preferences for the first condition, with ties counted as 0.5.
Full per-judge results are in Appendix~\ref{app:richness-full}.
}
\label{tab:richness-decomp}
\end{table*}

In the ASSET content-richness experiment, all five pairwise judges prefer the
meaning-equivalent richer clip. Because this manipulation
changes more than one surface property, Table~\ref{tab:richness-decomp}
separates content richness from duration and repetition controls. The
controls do not reproduce the preference: padding or repeating the concise clip
does not make it win, while richer speech remains preferred at matched duration.
The effect is therefore best interpreted as overweighting richer speech content,
rather than a simple preference for longer audio.

Pointwise scores show no consistent richer-speech advantage, and their
item-level relationship with pairwise preferences is weak
(Appendix~\ref{app:mode-divergence}). Human
listeners are also much less decisive: their mean richer-speech preference is
0.547 and its confidence interval crosses chance, compared with 0.707 for the
judges (Appendix~\ref{app:human-details}). The result is therefore not merely a shared preference for
longer utterances. Pairwise judges give content richness substantially more
weight than listeners do.

\keyfinding{Pairwise judges consistently favor content-rich speech, and the preference is much stronger than in human listeners.}

\subsection{Emotional Delivery Shapes Quality Judgments}
\label{sec:res-emotion}

\begin{figure}[t]
  \centering
  \includegraphics[width=.8\columnwidth]{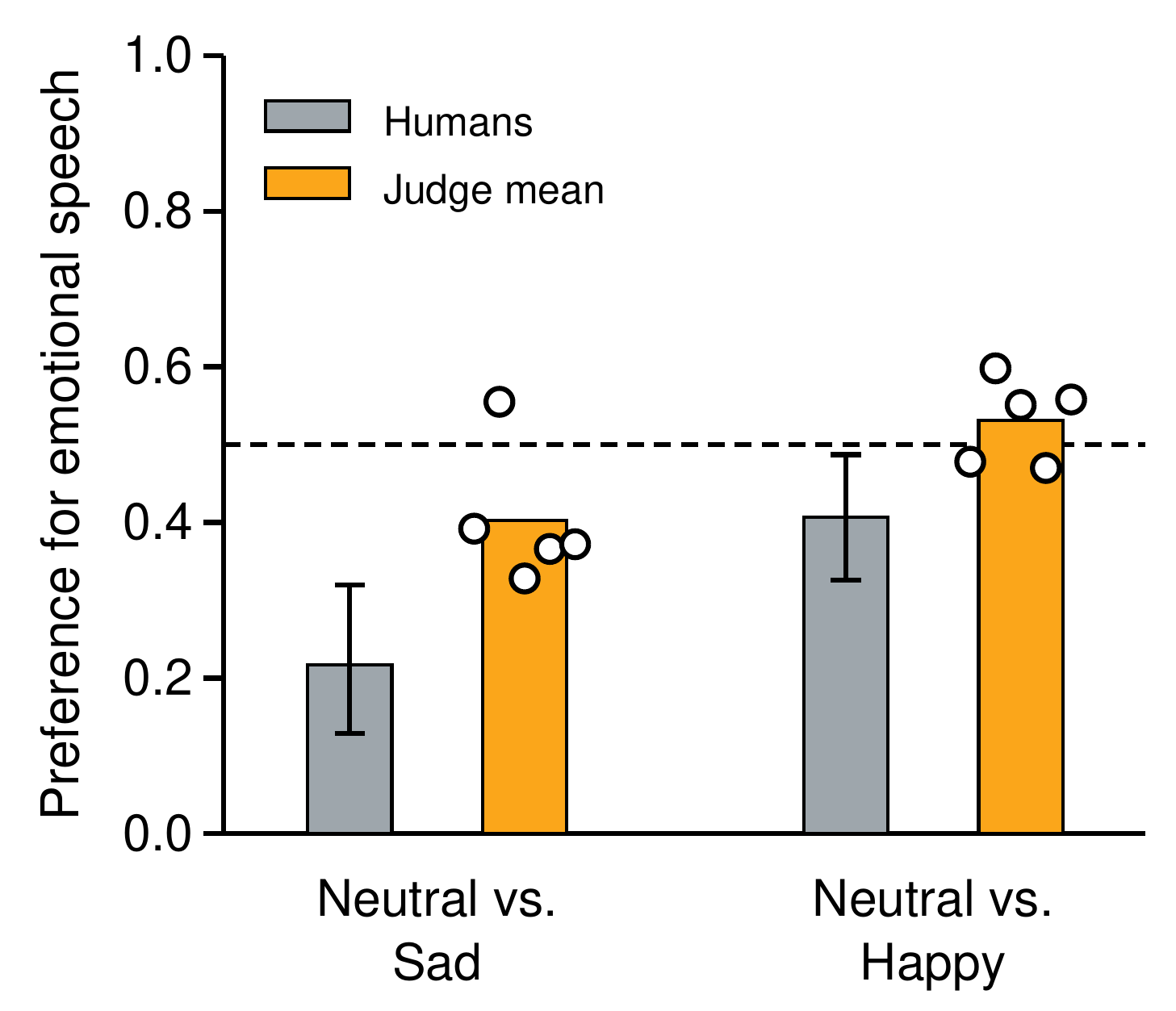}
  \caption{Human and mean judge preference for the emotional clip over the matched Neutral clip. White points show individual pairwise judges.}
  \label{fig:emotion-summary}
\end{figure}

\begin{figure*}[!t]
  \centering
  \includegraphics[width=\textwidth]{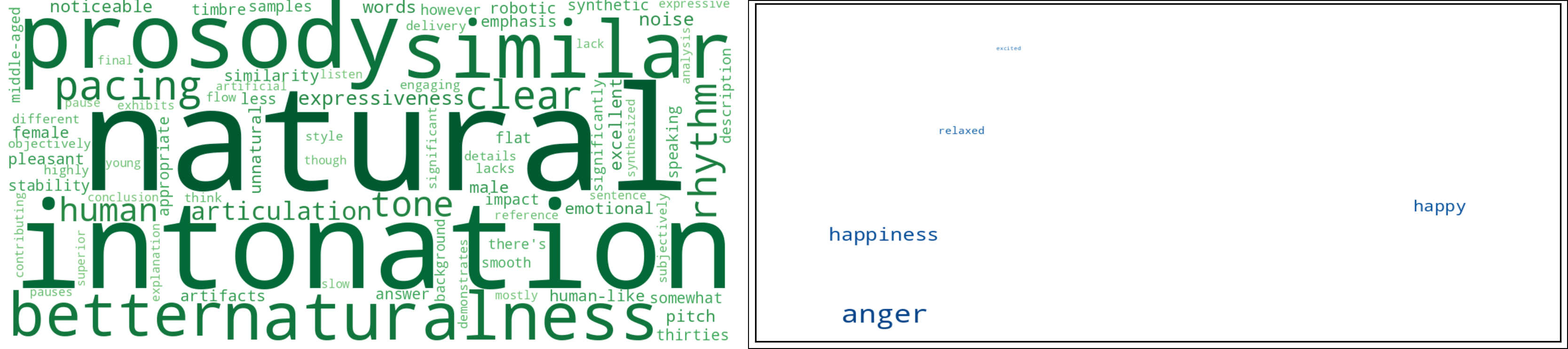}
  \caption{Rationale vocabulary for Neutral and emotional speech. Across
  acoustically different inputs, rationales repeatedly use similar quality and
  prosody terms rather than adapting to the specific emotional delivery.}
  \label{fig:rationale-wordclouds}
\end{figure*}

The ESD stimuli match speaker and semantic content across emotional conditions.
In pointwise scoring, four of five judges show a significant overall emotion
effect. The direction varies by judge: Sad and Angry are often scored lower than
Neutral, while Gemini-3.1-pro assigns higher scores to Happy and Angry.
In pairwise comparison, Sad loses to Neutral for four of five judges,
whereas Happy beats Neutral for UniSRM, SQ-LLM, and Gemini-3.1
(Appendix~\ref{app:emotion-full}).
Taken alone, the judge audit suggests that changes in intonation, speaking rate,
and expressiveness can shape quality preferences in both pointwise and pairwise
settings.

Human preference calibration changes how these effects should be interpreted.
Listeners strongly prefer Neutral over Sad: the Sad preference rate is 0.217,
compared with a judge mean of 0.403. The Sad penalty is therefore largely shared
with listeners rather than a judge-only shortcut. Happy shows a clearer
mismatch. The listener mean favors Neutral (Happy preference 0.407), while the
judge mean is 0.531 and three judges significantly favor Happy
(Figure~\ref{fig:emotion-summary}). 
Angry shows a weaker pattern: GRM and Gemini-2.5 prefer Neutral pairwise, while the other pairwise judges are near chance.
Emotional delivery can enter quality judgments, but the judge-specific
shortcut is clearer for Happy than for Sad.

\keyfinding{Human listeners and judges both penalize Sad speech, while Happy delivery shows the clearer judge-specific shortcut.}

\subsection{Pairwise Comparison Makes Acoustic Shortcuts More Visible}
\label{sec:diag-mode}

Across the three main experiments, pointwise scoring and pairwise comparison
are not interchangeable. Pairwise comparison consistently exposes the louder and
content-richness preferences, while pointwise scores are often weak or
resolution-limited. Emotional delivery affects both modes, but pairwise
comparison produces more consistent directional preferences across judges
(Appendix~\ref{app:mode-divergence}).

Boundary silence provides an additional diagnostic. UniSRM's pointwise scores
are nearly unchanged when silence is added, but its pairwise preference for the
padded clip falls to 0.069 when 1,000~ms of silence is added to each boundary.
Other judges show
mixed effects or many ties (Appendix~\ref{app:loudness-full}), so this is not a universal
silence-padding shortcut. It nevertheless illustrates how pairwise comparison can
lead a judge to act on an acoustic difference even when its pointwise score
barely moves.
Pairwise comparison is useful because it exposes small acoustic differences, but
the same sensitivity can make an invalid or overweighted cue more decisive in
the final judgment.

\keyfinding{Pairwise comparison exposes acoustic differences, but the same sensitivity can make invalid or overweighted cues more decisive.}

%% file: sections/05_rationales.tex
\section{Rationales Are Acoustically Underspecified}
\label{sec:rationales}
\label{sec:diag-cot}

Section~\ref{sec:results} shows that controlled acoustic changes can
systematically affect judge decisions. We next ask whether the accompanying
rationales name the acoustic evidence behind those changes. We separate two
problems. Section~\ref{sec:rationale-vocabulary} asks whether rationales adapt
to different acoustic inputs or fall back on the same quality vocabulary.
Section~\ref{sec:rationale-naming} asks a narrower question: when an acoustic
change affects the judgment, does the rationale say what changed?

\subsection{Rationales Draw on a Limited Recurring Vocabulary}
\label{sec:rationale-vocabulary}

Across manipulations, judge rationales repeatedly draw on a limited set of
descriptions. Words referring to naturalness, prosody, intonation, clarity, and
robotic speech remain common across conditions, while words specific to the
manipulation are much rarer. Automatic word-frequency analysis shows this
pattern for content richness, intensity, boundary silence, and emotional
delivery. Figure~\ref{fig:rationale-wordclouds} illustrates it for Neutral and
emotional speech.

This recurring vocabulary can describe the apparent quality of a clip without
fully contextualizing the acoustic evidence behind the judgment. For example,
when GRM strongly prefers Neutral over Sad delivery for the same speaker and
sentence, its rationale is one of the few to identify audible differences: the
Sad clip is described as having flatter prosody, slower pacing,
and more pauses, while also being called robotic. The rationale is therefore not
devoid of acoustic evidence. However, it treats these characteristics directly
as naturalness defects without first considering whether they express a
different intended emotion and then assessing naturalness within that delivery.
Appendix~\ref{app:cot-lexicons} provides the full example. This illustrates a
subtler grounding gap: a rationale may notice acoustic differences but fail to
contextualize them before converting them into quality penalties.

\keyfinding{Rationales repeatedly draw on broad quality vocabulary even when the acoustic input changes.}

\subsection{Rationales Rarely Name the Relevant Acoustic Change}
\label{sec:rationale-naming}

\begin{figure}[t]
  \centering
  \includegraphics[width=\columnwidth]{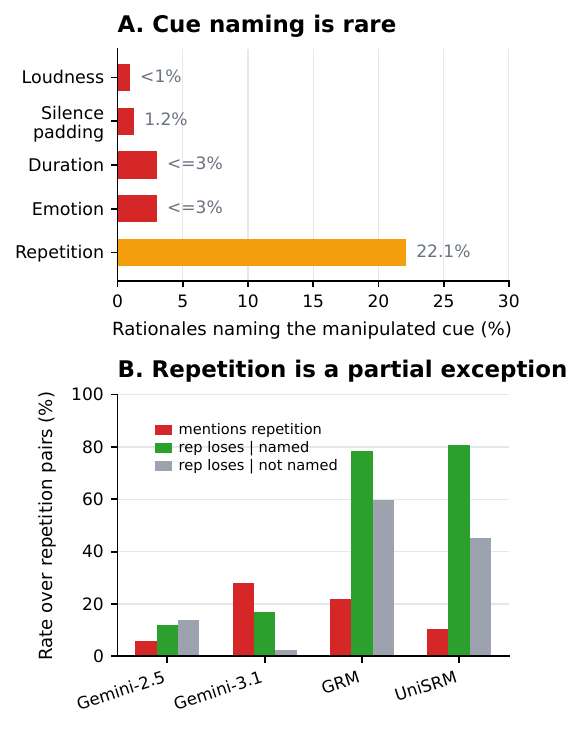}
  \caption{Rationale audit summary. Rationales rarely name the manipulated
  acoustic cue, while generic rubric terms remain common. Repetition is a
  partial exception: judges can sometimes name a conspicuous acoustic change,
  but do not do so reliably.}
  \label{fig:rationale-audit}
\end{figure}

Figure~\ref{fig:rationale-audit} shows the gap between changed judgments and
their explanations. Intensity changes affect judgments, yet direct
loudness-related terms such as ``loud,'' ``quiet,'' and ``higher volume'' are absent from
the audited UniSRM rationales. The same pattern appears
for boundary silence and content richness: the relevant words occur in at most
a few percent of rationales even when judge preferences change systematically.
For emotional delivery, judges rarely name the emotion that was
changed; in most cases, the rationale uses generic prosody or quality terms
instead of saying whether the clip sounds Happy, Sad, Angry, or Neutral.

As a duration control introduced in Section~\ref{sec:manipulations}, repetition
is a partial exception. Four judges mention repetition more often
than they name the other manipulations, with rates ranging from 5.7\% to
27.9\%. Even here, naming is incomplete, and mentioning repetition does
not consistently determine whether the repeated clip is penalized
(Appendix~\ref{app:cot-lexicons}). The visible rationales can therefore
sometimes name a conspicuous acoustic change, but do not do so reliably.

These results do not show that every rationale is factually incorrect or
unfaithful to the final answer. They show that visible explanations rely on a
limited recurring vocabulary and rarely name the acoustic evidence associated
with a systematic change in judgment. This limits the practical value of
rationales as debugging signals. If a judge
changes its decision with loudness, duration, or emotional delivery but
explains the decision using generic terms such as naturalness or prosody, the
rationale cannot tell users whether the judgment reflects the intended quality
or an acoustic shortcut.

\keyfinding{When an acoustic change affects a judgment, the rationale usually does not say what changed.}

%% file: sections/06_implications.tex
\section{Implications}
\label{sec:implications}

\paragraph{Controlled audits are needed to locate the failure mode.}
Our supporting analyses show what did not reliably change judgments. Within
our stimulus pools, we do not find stable preferences by speaker gender, speaker
accent, or Qwen-family self-preference, and broad low-level descriptors do not
explain the main effects (Appendix~\ref{app:natural-variance}). This does not
provide fairness guarantees, but it helps bound the claim: not every salient
attribute becomes a shortcut. Rather, specific acoustic cues become shortcuts
when the evaluation protocol makes them salient or lets judges give them too
much weight.

\paragraph{Using speech judges requires acoustic preprocessing and mode checks.}
Before using speech LLM judges for system comparison, data selection, or reward
modeling, evaluators should normalize loudness and trim leading or trailing
silence unless those cues are part of the target quality criterion. Pairwise
comparison should also be order-counterbalanced and reported alongside
pointwise scoring when both modes are available. Pairwise comparison is useful
because it exposes preferences that pointwise scores can hide, but the same
contrast can make invalid or overweighted cues more decisive. Human preference
calibration is especially important when a cue, such as content richness or
emotional delivery, may legitimately affect perceived quality.

\paragraph{Training better judges requires grounded rationale data.}
The rationale results suggest that simply asking for an explanation is not
enough. If supervised fine-tuning data mostly teaches rubric language such as
``naturalness,'' ``prosody,'' or ``expressiveness,'' a judge can produce fluent
rationales without identifying the acoustic evidence behind a changed decision.
Future speech-judge data should therefore include localized acoustic evidence:
the word, phrase, pronunciation, pause, loudness change, emotional delivery, or
other audible cue that justifies the judgment. It should also cover diverse
speakers, recording conditions, emotions, and synthesis systems so that judges
cannot rely on a narrow set of acoustic shortcuts.

\paragraph{Reasoning-oriented training is promising, but should be audited.}
Recent work shows that reinforcement learning can elicit or improve reasoning
behavior in language and speech models~\citep{deepseekr12025,speechjudge2025}.
For speech judges, this suggests a useful direction: training objectives could
reward rationales that connect scores to localized acoustic evidence, not only
rationales that are fluent or internally consistent with the final preference.
However, our findings imply that such training should itself be audited. A
rationale can be consistent with the final judgment while still failing to name
the acoustic cue that changed the judgment.

%% file: sections/09_conclusion.tex
\section*{Conclusion}

We audited six speech LLM judges with controlled acoustic manipulations and
human preference calibration. The results show that judges can turn perceptually
salient but invalid or overweighted cues into evidence of quality. They reward
louder audio even when only the signal level changes, prefer content-rich speech
more strongly than listeners do, and map emotional delivery into quality
preferences. These effects are clearest in pairwise comparison, which is useful
for exposing small acoustic differences but can also make acoustic shortcuts
more decisive.

More concerningly, the accompanying rationales do not reliably make these
decisions transparent.
Even when an acoustic manipulation changes a judgment, rationales rarely name
that change; instead, they often rely on broad rubric terms such as
``naturalness'' and ``prosody,'' leaving them acoustically underspecified.
Reliable speech judges therefore need two properties: their judgments should be
robust to acoustic shortcuts, and their rationales should connect decisions to
the acoustic evidence behind them.

%% file: sections/08_limitations.tex
\section{Limitations}
\label{sec:limitations}

This audit is controlled by design. All stimuli are English; the TTS pools cover
four systems and VCTK reference speakers; and the emotion pool uses ESD
recordings. These choices let us isolate specific cues, but they do not cover all
languages, accents, recording conditions, speakers, or modern TTS systems. Null
results for gender, accent, or low-level acoustic descriptors should therefore be
read as within-pool nulls, not broad fairness guarantees.

The judges also target different quality criteria. BTRM and GRM focus on naturalness,
whereas UniSRM, SQ-LLM, and Gemini prompts target broader overall quality. We
define bias relative to the intended quality criterion, but content richness and
emotional delivery may reasonably affect perceived quality. This is why we
calibrate them against human preference judgments rather than treating every
shift as bias by definition.

The human preference study is a calibration, not a full subjective benchmark. It covers
only the headline conditions and is large enough to separate the main
judge--listener mismatches, but not to estimate fine-grained human preferences
across many languages, speakers, emotions, and synthesis domains.

Finally, the rationale audit is lexical. A judge may internally represent a cue
without naming it, and rubric words such as ``prosody'' can sometimes refer to
multiple acoustic properties. Our claim is therefore limited to user-visible
rationales: they are acoustically underspecified because they do not reliably
and explicitly identify the manipulated acoustic change. We do not infer
from this omission that the cue is absent from the judge's internal representation.

%% file: sections/appendix.tex
\section*{Appendix}
\label{app:appendix}

\appsubsection{A.1}{TTS Systems}{app:tts}

Across TTS-generated experiments, we synthesize stimuli with four systems chosen
to span autoregressive (AR) and non-autoregressive (non-AR) generation, as well
as codec-token and mel-spectrogram output representations (Table~\ref{tab:tts}).
The goal is not to compare TTS systems, but to avoid tying the audit to one
synthesis architecture.

\input{tables/table_tts}

\appsubsection{A.2}{Stimulus Pools}{app:pools}

\paragraph{Acoustic manipulation pool.}
\label{app:pool-libritts}
The loudness and silence experiments use 80 base utterances filtered from
LibriTTS test-clean~\citep{libritts2019}. The text filter keeps short
declarative sentences, removes questions, exclamations, quotes, abbreviations,
loaded tokens, all-caps text, and mid-sentence proper nouns, and preserves a
spread over sentence lengths. Each utterance is synthesized once by one of four
TTS systems with a VCTK reference speaker~\citep{vctk2019}; the 10 reference
speakers are gender-balanced (40 female and 40 male assignments; speaker IDs
p225, p227, p256, p267, p276, p294, p301, p311, p345, and p363). Intensity
adjustments and silence padding are post-hoc waveform transforms on the same base
waveform. The silence conditions add 0.5 or 1 second at both the beginning and
end of each utterance.

\paragraph{Content-richness pool.}
\label{app:pool-asset}
The content-richness experiment uses ASSET sentence-simplification pairs
~\citep{asset2020}. The concise side is a human simplification and the richer
side is the original meaning-equivalent Wikipedia sentence. After excluding one
catastrophic TTS failure, the analysis uses 149 pairs, approximately balanced
over TTS systems (38/37/37/37) and speaker gender (75 female, 74 male). Each
concise/richer pair is synthesized with the same TTS system, speaker, and
reference voice. One representative pair is:
\begin{quote}
\textbf{Concise:} Perth is used for other settlements worldwide.

\textbf{Richer:} The name Perth has hence been used for a number of other
settlement around the world.
\end{quote}
Silence-padding and repetition controls are derived from the concise side of the
same pair.

\paragraph{Emotion pool.}
\label{app:pool-esd}
The emotion experiment uses English ESD recordings~\citep{esd2022}. We use 10
English speakers, 20 texts, and four emotions: Neutral, Happy, Sad, and Angry.
All clips are converted to a common sampling format, VAD-trimmed, and normalized
to -23 LUFS before scoring. Pairwise emotion trials compare each emotional clip
against the matched Neutral clip from the same speaker and text, in both A/B
orders.

\appsubsection{A.3}{Prompts and Reproducibility}{app:reproducibility}

\paragraph{Gemini pairwise prompt.}
\label{app:gemini-prompt}
For the Gemini pairwise experiments, we use a concatenated-audio setup following
AudioJudge-style test-concat evaluation. The system prompt tells the model that
the audio contains two speech clips separated by one second of silence and asks
it to compare clarity, naturalness, and overall quality. The response is
constrained to JSON with a reasoning field and a label:

\newpage
\begin{promptbox}{Gemini Pairwise Speech-Quality Prompt}
\small
You are an expert in audio quality assessment specializing in synthesized speech
evaluation. The audio you will receive contains TWO speech clips concatenated
with 1 second of silence between them: Audio 1 followed by 1 second of silence,
then Audio 2. Your task is to critically compare Audio 1 and Audio 2 on clarity,
naturalness, and overall quality. \ldots{} Respond only with valid JSON with
keys \texttt{reasoning} and \texttt{label}, where \texttt{label} is
\texttt{"1"}, \texttt{"2"}, or \texttt{"tie"}.
\end{promptbox}

The position/length ablation appends the following line to the system prompt:

\begin{promptbox}[colback=promptredbg,colframe=promptredframe]{Position/Length Prompt Line}
\small
Avoid position bias and don't let response length influence your evaluation.
\end{promptbox}

\paragraph{Other judge prompts.}
\label{app:other-prompts}
For GRM, we use a target-free naturalness-only variant for the content-richness
experiment. The released GRM prompt normally includes target text; in the
concise/richer setting this would conflate speech quality with text fidelity.
For UniSRM, we use its released pointwise prompt for single-clip scoring and a
quality-only pairwise prompt for target-free comparisons. UniSRM's Task-AB style
prompt includes text-fidelity dimensions; those conditions are excluded from the
main content-richness analysis for the same reason as GRM target-text prompts.
SQ-LLM is run with its SingleEval and CompareEval formats.

\paragraph{Released audit suite.}
We release the controlled stimuli, manifests, prompt templates, and analysis
scripts used in this study as \textsc{SpeechJudgeAudit}. The suite allows the
same tests to be applied to additional speech judges, including tests of
rationale grounding when rationales are available. For ESD, we provide the
selected file list and a preparation script rather than redistributing the
original recordings.

\appsubsection{B.1}{Statistical Reporting}{app:stats}

For pairwise experiments, the reported effect is the manipulated-clip preference
rate, with manipulated-clip wins scored as 1, reference-clip wins as 0, and ties
as 0.5. When both A/B and B/A orders are available, we first aggregate the two
orders for the same underlying item and then aggregate across items. For pointwise experiments, the
reported effect is the within-judge manipulated--reference score difference.
Confidence intervals are item-level bootstraps unless otherwise stated. In the
full tables, ``degenerate'' denotes zero variance or integer-scale collapse, and
``low-power'' marks high-tie pairwise settings with too few decisive outcomes.
Significance markers are * ($p<.05$), ** ($p<.01$), and *** ($p<.001$);
``n.s.'' denotes a non-significant result.

\appsubsection{B.2}{Loudness and Silence Results}{app:loudness-full}
\label{app:silence-full}

Table~\ref{tab:loudness-combined} reports loudness effects side by side:
every pairwise judge prefers $+6$ dB, while pointwise often hides it.
Table~\ref{tab:silence-combined} shows the analogous silence result: UniSRM
pointwise is nearly unchanged, but pairwise rejects padding.

\begin{table*}[t]
\centering
\scriptsize
\setlength{\tabcolsep}{3pt}
\input{generated/loudness_combined}
\caption{Combined loudness results. Pairwise preference is
$P(\mathrm{manipulated} > \mathrm{reference})$ with ties counted as 0.5; pointwise
$\Delta$ is the within-judge manipulated--reference score difference. All pairwise
judges prefer $+6$ dB clips, while pointwise scoring often hides this
preference.}
\label{tab:loudness-combined}
\end{table*}

\begin{table*}[t]
\centering
\scriptsize
\setlength{\tabcolsep}{3pt}
\input{generated/silence_combined}
\caption{Combined silence-padding results. Pointwise $\Delta$ is the padded--base
score difference; pairwise pad preference is $P(\mathrm{padded} >
\mathrm{base})$ with ties counted as 0.5. UniSRM illustrates the diagnostic
pattern most clearly: pointwise scores are nearly unchanged, but pairwise
comparison strongly rejects padding.}
\label{tab:silence-combined}
\end{table*}

\appsubsection{B.3}{Content-Richness Results}{app:richness-full}

Table~\ref{tab:gemini-ablation} shows that an explicit position/length
instruction does not
remove Gemini-3.1's content-richness preference. Table~\ref{tab:richness-full}
shows that richer speech is preferred by every pairwise judge, while silence-padding
and repetition do not reproduce the effect. Text-fidelity controls are needed
because target transcripts can confound speech quality with transcript matching
(Table~\ref{tab:text-fidelity-controls}).

\begin{table}[t]
\centering
\small
\setlength{\tabcolsep}{3pt}
\input{generated/prompt_ablation_gemini}
\caption{Gemini-3.1 position/length prompt ablation on ASSET content-richness pairs.
The explicit instruction has negligible effect: the richer-speech preference remains near
0.70.}
\label{tab:gemini-ablation}
\end{table}

\begin{table*}[t]
\centering
\small
\setlength{\tabcolsep}{3pt}
\input{generated/richness_decomposition_stats}
\caption{Full content-richness decomposition. Values are manipulated-clip preference
rates with ties counted as 0.5. Richer denotes the original ASSET sentence, C =
concise, Pad = silence-padded concise, and Rep = repeated concise; significance markers follow
Appendix~\ref{app:stats}.}
\label{tab:richness-full}
\end{table*}

\begin{table*}[t]
\centering
\scriptsize
\setlength{\tabcolsep}{3pt}
\input{generated/text_fidelity_controls}
\caption{Text-fidelity and prompt controls for the content-richness experiment.
Target-text conditions can confound acoustic preference by rewarding or
penalizing extra words through transcript matching.}
\label{tab:text-fidelity-controls}
\end{table*}

\appsubsection{B.4}{Emotional Delivery Results}{app:emotion-full}

\begin{table*}[t]
\centering
\scriptsize
\setlength{\tabcolsep}{3pt}
\input{generated/emotion_combined}
\caption{Combined emotion results. Pairwise preference is
$P(\mathrm{emotion} > \mathrm{Neutral})$ with ties counted as 0.5; pointwise
$\beta$ is the z-scored coefficient relative to Neutral. Overall significance is
from the likelihood-ratio test for the emotion fixed effects.}
\label{tab:emotion-combined}
\end{table*}

The emotion analysis first asks whether pointwise scores respond to emotional
delivery at all. For each judge, we z-score scores within judge and fit a
mixed-effects model with emotion as a fixed effect and random intercepts for
speaker and text:
\[
  \begin{aligned}
  \texttt{score\_z} &\sim \texttt{emotion} \\
  &\quad + (1 \mid \texttt{speaker}) + (1 \mid \texttt{text}) .
  \end{aligned}
\]
Here \texttt{score\_z} is the judge score z-scored within judge;
\texttt{emotion} uses Neutral as the reference level; and the two random
intercepts account for repeated speakers and repeated texts. The pointwise
$\beta$ values in Table~\ref{tab:emotion-combined} are the estimated emotion
contrasts against Neutral. The overall significance column comes from a
likelihood-ratio test comparing this model against one without the emotion fixed
effect. Four of the five pointwise judges show a significant overall emotion
effect, indicating that their scores are sensitive to emotional delivery.

The second question is whether this sensitivity becomes a directional quality
preference under pairwise comparison. Table~\ref{tab:emotion-combined} compares
each emotional clip against the matched Neutral clip from the same speaker and
text. Sad loses to Neutral for four of five pairwise judges, with UniSRM as the
exception. Happy is above chance for UniSRM, SQ-LLM, and Gemini-3.1, but flat
for GRM and Gemini-2.5. Thus judges do not merely perceive emotional variation;
several map it into quality preferences.

\appsubsection{B.5}{Pointwise and Pairwise Comparison}{app:mode-divergence}

Table~\ref{tab:mode-divergence} summarizes where pairwise comparison reveals
directional preferences that pointwise scores obscure.

\begin{table*}[t]
\centering
\scriptsize
\setlength{\tabcolsep}{4pt}
\input{generated/mode_divergence_details}
\caption{Pointwise and pairwise evidence across controlled acoustic
manipulations. Pairwise comparison often reveals a directional preference that
is weak, noisy, or absent in pointwise scores.}
\label{tab:mode-divergence}
\end{table*}

\appsubsection{B.6}{Position-Bias Diagnostics}{app:position-bias}

Pairwise trials are order-counterbalanced because several judges show strong
presentation-order effects. These diagnostics are not treated as acoustic
shortcuts, but they are important for interpreting pairwise preferences:
without A/B and B/A averaging, a position effect could be mistaken for an
acoustic preference.

\begin{table*}[t]
\centering
\small
\begin{tabularx}{\textwidth}{p{0.20\textwidth} Y Y}
\toprule
Setting & Position effect observed & Role in this audit \\
\midrule
Content richness & GRM strongly favors the second clip in decisive comparisons
(B wins 91.6\%); Gemini-2.5 also favors B (71.8\% of trials), while UniSRM
and SQ-LLM lean toward A (64.8\% and 56.4\%). & We average the two presentation
orders before computing the richer-speech preference, so the reported effect is
not a single-order artifact. \\
Loudness & GRM shows order asymmetries of roughly 10--15 percentage points
across loudness conditions even when its order-averaged loudness preference is
near chance. & Order averaging prevents a positional preference from being
interpreted as evidence for or against a loudness preference. \\
Silence padding & GRM and SQ-LLM show clear AB/BA differences, while their
order-averaged padded-clip preferences remain near chance. & This supports
using order-averaged item preferences and treating the strongest silence result
as judge-specific rather than universal. \\
Emotional delivery & The average AB--BA gap is about 0.10 for SQ-LLM, 0.48 for
GRM, 0.61 for Gemini-2.5, 0.68 for Gemini-3.1, and 0.80 for UniSRM. & The
emotion preferences in Table~\ref{tab:emotion-combined} are interpretable
because they are averaged across both orders. \\
\bottomrule
\end{tabularx}
\caption{Position-bias diagnostics for pairwise experiments. The numbers are
used to justify order counterbalancing, not as acoustic-shortcut results.}
\label{tab:position-bias}
\end{table*}

\appsubsection{C.1}{Rationale Audits}{app:cot-lexicons}

The rationale audit separates behavior from explanation. The key pattern is that
rationales often use generic rubric language while failing to name the
manipulated acoustic cue. A representative pairwise example is:

\newpage
\begin{promptbox}{Rationale example: Sad vs. Neutral}
\small
\textbf{GRM pairwise output, excerpted verbatim.} ESD speaker 0012 reads
``One two three, where is your breakfast'' in Neutral (A) and Sad (B) delivery.

\textbf{Output A:} ``The intonation is very natural and human-like... The pacing
is smooth and consistent, mirroring natural speech... This output sounds
remarkably natural, almost indistinguishable from a human speaker.''

\textbf{Output B:} ``The prosody is quite unnatural... very flat and robotic...
The pacing is slow and deliberate, with noticeable and unnatural pauses... This
output sounds highly artificial and robotic.''

\textbf{Conclusion:} Output A: 9; Output B: 2.
\end{promptbox}

The rationale identifies several audible differences, but treats them as
naturalness defects without contextualizing them as Sad delivery before
assigning the lower score.

Figure~\ref{fig:rationale-audit} summarizes the same pattern across
manipulations. Terms naming the manipulated acoustic change are rare for loudness, silence, duration, and
emotion. Repetition is the partial exception: some judges can name it, but
detection remains incomplete and does not fully determine the penalty.

The explicit-cue lexicons are narrow by design. For intensity, we count direct
references to loudness, volume, quietness, amplitude, level, or dB. For boundary
silence, we count silence, padding, leading/trailing silence, pause, and gap
terms when they refer to the clip boundary. For content richness, we count
length, verbosity, richer content, extra words, and related wording. For
emotional delivery, we count emotion-specific words for Happy, Sad, Angry, and
Neutral, while excluding generic prosody terms such as ``tone,'' ``intonation,''
``prosody,'' and ``expressive.'' For repetition, we count repeat, repetition,
duplicate, and related terms.
\begin{table*}[t]
\centering
\small
\input{generated/cot_cause_naming_summary}
\caption{Summary of explicit acoustic-term rates in rationales. Generic rubric
terms refer to broad quality or prosody language rather than the manipulated
acoustic change.}
\label{tab:cot-cause-summary}
\end{table*}

\begin{table*}[t]
\centering
\small
\input{generated/repetition_detection_penalty}
\caption{Repetition detection versus penalty. Mentions rep. is the fraction of
rep-vs-concise rationales mentioning repetition keywords; the conditional
columns report how often the repeated clip loses when repetition is or is not
named.}
\label{tab:rep-detect-penalize}
\end{table*}

\appsubsection{D.1}{Human Preference Calibration}{app:human-details}

\begin{table*}[t]
\centering
\scriptsize
\setlength{\tabcolsep}{3.6pt}
\input{generated/human_mos_combined}
\caption{Human preference calibration and agreement diagnostics. Human mean is the
item-level manipulated-clip preference rate with bootstrap 95\% confidence intervals.
Human significance uses a two-sided sign-flip test against 0.5. Judge mean is
averaged over the five pairwise judges, the bracketed range shows their spread,
and Sig. judges counts how many individual judges are significant in either
direction.}
\label{tab:human-mos-combined}
\end{table*}

\begin{table*}[t]
\centering
\scriptsize
\setlength{\tabcolsep}{4pt}
\input{generated/human_rater_qc}
\caption{Human listener quality-control diagnostics. Majority agreement is the
fraction of a rater's responses that match the item-level majority label;
item-mean MAE is the mean absolute error between the rater's numeric response
and the item mean after coding manipulated/reference/tie as 1/0/0.5.}
\label{tab:human-rater-qc}
\end{table*}

The human preference calibration uses 15 listeners with native or near-native English
proficiency. Each listener was assigned 30 pairs and responded with A, B, or tie,
yielding 449 valid judgments. The 60 unique pairs comprise 12 loudness pairs, 18 content-richness pairs, 15 Sad-vs-Neutral
pairs, and 15 Happy-vs-Neutral pairs. The instruction is: ``These clips are
synthesized or recorded speech samples. For each pair, choose which utterance
has better overall naturalness and speech quality.'' Table~\ref{tab:human-mos-combined}
combines manipulated-clip preference rates, human significance tests, judge residuals,
and agreement diagnostics. Entropy is computed over the item-level distribution
of \{manipulated, tie, reference\}; higher entropy and variance indicate more
listener disagreement. Human brackets are bootstrap confidence intervals over
items. Judge brackets are ranges over the five pairwise judges, so they describe
model spread rather than sampling uncertainty. Table~\ref{tab:human-rater-qc}
reports listener-level quality-control diagnostics.

Table~\ref{tab:human-mos-combined} supports three interpretations used in the
main text. Higher loudness is perceptually salient for humans, but remains
invalid evidence of better TTS quality for a judge. Content richness is near
chance for humans but consistently preferred by
judges. For emotion, Sad-vs-Neutral has strong listener agreement in favor of
Neutral, whereas Happy-vs-Neutral is more split for humans while three of five
judges significantly favor Happy.

\appsubsection{E.1}{Supporting Analyses}{app:natural-variance}

Table~\ref{tab:natural-variance} reports supporting diagnostics for the scope of
the audit. They show which attributes did not reliably change judgments within
our pools, and which associations remain diagnostic rather than causal.

\begin{table*}[t]
\centering
\small
\input{generated/natural_variance_summary}
\caption{Natural-variance and supporting analyses. These diagnostics are
within-pool evidence and should not be read as causal mechanisms or broad
fairness guarantees.}
\label{tab:natural-variance}
\end{table*}
\FloatBarrier

%% file: tables/table_tts.tex
\begin{table}[b]
\centering
\footnotesize
\begin{tabularx}{\columnwidth}{l Y Y}
\toprule
\textbf{System} & \textbf{Generation structure} & \textbf{Representation} \\
\midrule
CosyVoice2 & AR LLM + flow matching & Codec tokens \\
Qwen3-TTS & AR LLM + codec decoder & Codec tokens \\
F5-TTS & Non-AR DiT flow matching & Mel-spectrogram \\
StyleTTS2 & Non-AR diffusion + GAN & Mel-spectrogram \\
\bottomrule
\end{tabularx}
\caption{TTS systems used for generated-speech stimuli. They are chosen to span
autoregressive (AR) and non-autoregressive (non-AR) generation, as well as
codec-token and mel-spectrogram output representations.}
\label{tab:tts}
\end{table}

%% file: generated/loudness_combined.tex
% Combined loudness results. Pairwise effect = preference rate; pointwise effect = score delta.
\begin{tabularx}{\textwidth}{l l c c c c c c Y}
\toprule
Judge & Condition & Pairwise N & Pairwise pref. & Pairwise sig. & Pointwise N & Pointwise $\Delta$ & Pointwise sig. & Notes \\
\midrule
BTRM & $-6$ dB vs. ref & -- & -- & -- & 100 & -0.54 & ** & quiet penalty \\
BTRM & $-3$ dB vs. ref & -- & -- & -- & 99 & -0.42 & ** & quiet penalty \\
BTRM & $+3$ dB vs. ref & -- & -- & -- & 99 & +0.05 & n.s. &  \\
BTRM & $+6$ dB vs. ref & -- & -- & -- & 99 & -0.12 & n.s. &  \\
\midrule
GRM & $-6$ dB vs. ref & 80 & 0.500 & n.s. & -- & -- & -- & pairwise-only judge \\
GRM & $-3$ dB vs. ref & 80 & 0.475 & n.s. & -- & -- & -- &  \\
GRM & $+3$ dB vs. ref & 80 & 0.516 & n.s. & -- & -- & -- &  \\
GRM & $+6$ dB vs. ref & 80 & 0.541 & * & -- & -- & -- & louder wins \\
\midrule
UniSRM & $-6$ dB vs. ref & 80 & 0.469 & n.s. & 159 & +0.04 & n.s. &  \\
UniSRM & $-3$ dB vs. ref & 80 & 0.481 & n.s. & 159 & +0.01 & n.s. &  \\
UniSRM & $+3$ dB vs. ref & 80 & 0.512 & n.s. & 159 & +0.04 & n.s. &  \\
UniSRM & $+6$ dB vs. ref & 80 & 0.575 & ** & 159 & +0.09 & ** & both methods agree \\
\midrule
SQ-LLM & $-6$ dB vs. ref & 80 & 0.472 & n.s. & 160 & +0.01 & n.s. & high pairwise ties \\
SQ-LLM & $-3$ dB vs. ref & 80 & 0.484 & n.s. & 160 & +0.01 & n.s. & high pairwise ties \\
SQ-LLM & $+3$ dB vs. ref & 80 & 0.506 & n.s. & 160 & 0.00 & deg. & integer-scale collapse \\
SQ-LLM & $+6$ dB vs. ref & 80 & 0.591 & *** & 160 & 0.00 & deg. & pairwise reveals preference \\
\midrule
Gemini-2.5 & $-6$ dB vs. ref & 80 & 0.431 & *** & 156 & -0.09 & n.s. & ref wins pairwise \\
Gemini-2.5 & $-3$ dB vs. ref & 80 & 0.497 & n.s. & 155 & -0.11 & n.s. &  \\
Gemini-2.5 & $+3$ dB vs. ref & 80 & 0.503 & n.s. & 155 & -0.23 & n.s. &  \\
Gemini-2.5 & $+6$ dB vs. ref & 80 & 0.566 & *** & 155 & -0.13 & n.s. & pairwise reveals direction \\
\midrule
Gemini-3.1 & $-6$ dB vs. ref & 80 & 0.384 & *** & 155 & -0.15 & n.s. & monotonic pairwise pattern \\
Gemini-3.1 & $-3$ dB vs. ref & 80 & 0.481 & n.s. & 155 & -0.16 & n.s. &  \\
Gemini-3.1 & $+3$ dB vs. ref & 80 & 0.528 & * & 155 & -0.09 & n.s. & louder wins pairwise \\
Gemini-3.1 & $+6$ dB vs. ref & 80 & 0.616 & *** & 154 & -0.19 & n.s. & strongest pairwise effect \\
\bottomrule
\end{tabularx}

%% file: generated/silence_combined.tex
% Combined silence-padding results. Pairwise effect = padded preference rate; pointwise effect = padded - base score delta.
\begin{tabularx}{\textwidth}{l l c c c c c c Y}
\toprule
Judge & Condition & Pairwise N & Pairwise pref. & Pairwise sig. & Pointwise N & Pointwise $\Delta$ & Pointwise sig. & Notes \\
\midrule
BTRM & pad500 vs. base & -- & -- & -- & 80 & -1.59 & * & pad penalty \\
BTRM & pad1000 vs. base & -- & -- & -- & 80 & -2.87 & *** & pad penalty grows \\
\midrule
UniSRM & pad500 vs. base & 80 & 0.250 & *** & 80 & -0.01 & n.s. & pointwise 84\% identical \\
UniSRM & pad1000 vs. base & 80 & 0.069 & *** & 80 & +0.01 & n.s. & pointwise 86\% identical \\
\midrule
SQ-LLM & pad500 vs. base & 80 & 0.537 & n.s. & 80 & +0.01 & n.s. & 96\% identical scores \\
SQ-LLM & pad1000 vs. base & 80 & 0.422 & n.s. & 80 & -0.04 & n.s. & trend anti-pad \\
\midrule
Gemini-2.5 & pad500 vs. base & 80 & 0.512 & n.s. & 80 & +0.59 & ** & pointwise pad preference \\
Gemini-2.5 & pad1000 vs. base & 80 & 0.506 & n.s. & 80 & +0.49 & * & pairwise low-power \\
\midrule
Gemini-3.1 & pad500 vs. base & 80 & 0.503 & n.s. & 80 & -0.24 & n.s. & pairwise low-power \\
Gemini-3.1 & pad1000 vs. base & 80 & 0.509 & n.s. & 80 & -0.32 & n.s. & pairwise low-power \\
\bottomrule
\end{tabularx}

%% file: generated/prompt_ablation_gemini.tex
% Gemini-3.1 position/length prompt ablation summary.
\begin{tabularx}{\columnwidth}{l c Y}
\toprule
Setting & Richer preference & Takeaway \\
\midrule
Baseline & 0.701 & reference \\
Position/length instruction & 0.695 & preference remains \\
\bottomrule
\end{tabularx}

%% file: generated/richness_decomposition_stats.tex
% Content-richness decomposition. Values are manipulated-clip preference rates; significance uses sign-flip test vs 0.5.
\begin{tabularx}{\textwidth}{l c c c c c c c c}
\toprule
Judge & Richer $>$ concise & Sig. & Pad $>$ concise & Sig. & Repeat $>$ concise & Sig. & Richer $>$ pad & Sig. \\
\midrule
UniSRM & 0.846 & *** & 0.082 & *** & 0.447 & ** & 0.954 & *** \\
GRM & 0.570 & *** & 0.527 & n.s. & 0.362 & *** & 0.567 & *** \\
SQ-LLM & 0.755 & *** & 0.424 & * & 0.515 & n.s. & 0.765 & *** \\
Gemini-2.5 & 0.664 & *** & 0.507 & n.s. & 0.515 & n.s. & 0.544 & ** \\
Gemini-3.1 & 0.701 & *** & 0.502 & n.s. & 0.471 & *** & 0.661 & *** \\
\bottomrule
\end{tabularx}

%% file: generated/text_fidelity_controls.tex
% Target-text / text-fidelity controls for content-richness experiment.
\begin{tabularx}{\textwidth}{l Y c c Y}
\toprule
Judge & Setting & Text fidelity? & Richer pref. & What it shows \\
\midrule
BTRM & Concise target transcript supplied & yes & 0.705 & target transcript changes the outcome \\
BTRM & Richer target transcript supplied & yes & 0.262 & target mismatch reverses the direction \\
BTRM & Concise and richer targets averaged & yes & 0.483 & target effects cancel; no target-free analogue \\
UniSRM & Target transcript plus text-fidelity dimension & yes & -- & text fidelity masks content-richness preference \\
UniSRM & Target-free prompt variant & no & -- & output became template-like, so excluded \\
Gemini-3.1 & Anti-position/length instruction only & no & 0.695 & preference remains despite the instruction \\
\bottomrule
\end{tabularx}

%% file: generated/emotion_combined.tex
% Combined emotion results. Pairwise preference = P(emotion > Neutral); pointwise beta uses Neutral baseline.
\begin{tabularx}{\textwidth}{l l c c c c c c Y}
\toprule
Judge & Emotion & Pairwise N & Pairwise pref. & Pairwise sig. & Pointwise $\beta$ & Pointwise sig. & Overall sig. & Notes \\
\midrule
BTRM & Happy & -- & -- & -- & -0.085 & n.s. & ** &  \\
BTRM & Sad & -- & -- & -- & -0.205 & *** & ** & lower than Neutral \\
BTRM & Angry & -- & -- & -- & -0.129 & * & ** & lower than Neutral \\
\midrule
GRM & Happy & 400 & 0.478 & n.s. & -- & -- & -- & flat \\
GRM & Sad & 400 & 0.392 & *** & -- & -- & -- & Neutral wins \\
GRM & Angry & 400 & 0.395 & *** & -- & -- & -- & Neutral wins \\
\midrule
UniSRM & Happy & 400 & 0.598 & *** & -0.040 & n.s. & *** & Happy wins pairwise \\
UniSRM & Sad & 400 & 0.555 & * & +0.072 & n.s. & *** & pro-Sad outlier \\
UniSRM & Angry & 400 & 0.535 & n.s. & -0.392 & *** & *** & pointwise penalty \\
\midrule
SQ-LLM & Happy & 400 & 0.551 & * & -0.126 & n.s. & *** & Happy wins pairwise \\
SQ-LLM & Sad & 400 & 0.328 & *** & -0.427 & *** & *** & strongest Sad penalty \\
SQ-LLM & Angry & 400 & 0.473 & n.s. & -0.142 & n.s. & *** & flat pairwise \\
\midrule
Gemini-2.5 & Happy & 400 & 0.470 & n.s. & -0.267 & ** & n.s. & mixed evidence \\
Gemini-2.5 & Sad & 400 & 0.366 & *** & -0.248 & ** & n.s. & Neutral wins \\
Gemini-2.5 & Angry & 400 & 0.429 & ** & -0.310 & *** & n.s. & Neutral wins \\
\midrule
Gemini-3.1 & Happy & 400 & 0.558 & * & +0.250 & ** & *** & Happy wins \\
Gemini-3.1 & Sad & 400 & 0.372 & *** & -0.272 & ** & *** & Neutral wins \\
Gemini-3.1 & Angry & 400 & 0.482 & n.s. & +0.188 & * & *** & pointwise positive \\
\bottomrule
\end{tabularx}

%% file: generated/mode_divergence_details.tex
% Compact support table for pointwise/pairwise divergence.
\begin{tabularx}{\textwidth}{p{0.18\textwidth} Y Y Y}
\toprule
Cue / example & Pointwise evidence & Pairwise evidence & Takeaway \\
\midrule
Content richness & no consistent richer-speech advantage; Gemini-2.5 leans concise & richer preference 0.570--0.846 & pairwise comparison reveals a consistent preference \\
Content richness correlations & Gemini-3.1 $r=+0.14$; SQ-LLM $r=+0.096$ & same items, pairwise preferences & per-item pointwise and pairwise verdicts are weakly related \\
Silence / UniSRM & $\Delta\approx0$, 84--86\% identical scores & pad1000 preference 0.069 & integer scale hides anti-pad preference \\
Loudness / SQ-LLM & loud6 $\Delta=0.000$, degenerate & loud6 preference 0.591 & pointwise resolution ceiling hides pro-loud preference \\
Loudness / Gemini & all pointwise deltas n.s. & quiet6 loses, loud6 wins & pointwise noise hides directional level preference \\
Emotional delivery & Kendall $W=0.33$ across pointwise ranks & Kendall $W=0.84$ across pairwise ranks & pairwise rankings are more consistent across judges \\
\bottomrule
\end{tabularx}

%% file: generated/cot_cause_naming_summary.tex
% Explicit acoustic-term summary used by diagnostics.
\begin{tabularx}{\textwidth}{p{0.20\textwidth} p{0.20\textwidth} Y Y}
\toprule
Manipulation & Judge(s) audited & Explicit acoustic-term rate & Generic rubric terms \\
\midrule
Content richness / length & Gemini-3.1, GRM, UniSRM & length 0--3\%; richer/original 0\% & rubric/prosody terms frequent \\
Loudness & UniSRM & loud/loudness/volume 0\%; amplitude/dB/level 0.94\% & prosody/naturalness/quality 100\% \\
Silence padding & UniSRM & silence 0\%; duration 0\%; pad/leading/trailing 1.25\% & prosody/naturalness/quality 100\% \\
Emotion & rationale-producing judges & emotion-specific terms rarely identify the true emotion (3/16 pointwise cells) & generic prosody terms excluded from lexicon \\
Repetition & GRM & repetition keywords 22.1\% for rep-vs-concise & partial exception; detection incomplete \\
\bottomrule
\end{tabularx}

%% file: generated/repetition_detection_penalty.tex
% Repetition detection vs penalty summary.
\begin{tabularx}{\textwidth}{lrrr}
\toprule
Judge & Mentions rep. & $P(\mathrm{rep\ loses}\mid\mathrm{mentioned})$ & $P(\mathrm{rep\ loses}\mid\mathrm{not})$ \\
\midrule
Gemini-2.5 & 5.7\% & 11.8\% & 13.9\% \\
Gemini-3.1 & 27.9\% & 16.9\% & 2.3\% \\
GRM & 21.8\% & 78.5\% & 59.7\% \\
UniSRM & 10.5\% & 80.6\% & 45.1\% \\
\bottomrule
\end{tabularx}

%% file: generated/human_mos_combined.tex
% Combined human preference calibration and agreement diagnostics.
\begin{tabularx}{\textwidth}{l c c c c c c c c c}
\toprule
Condition & Items & N & Human mean [CI] & Human sig. & Judge mean [range] & Sig. judges & Residual & Human entropy & Human variance \\
\midrule
+6 dB vs. ref & 12 & 88 & 0.612 [0.548, 0.671] & * & 0.578 [0.541, 0.616] & 5/5 & -0.035 & 1.009 & 0.071 \\
Richer vs. concise & 18 & 134 & 0.547 [0.464, 0.627] & n.s. & 0.707 [0.570, 0.846] & 5/5 & +0.160 & 1.260 & 0.111 \\
Sad vs. Neutral & 15 & 115 & 0.217 [0.129, 0.320] & *** & 0.403 [0.328, 0.555] & 5/5 & +0.186 & 0.756 & 0.098 \\
Happy vs. Neutral & 15 & 112 & 0.407 [0.326, 0.488] & n.s. & 0.531 [0.470, 0.598] & 3/5 & +0.124 & 1.305 & 0.155 \\
\bottomrule
\end{tabularx}

%% file: generated/human_rater_qc.tex
% Human rater QC diagnostics.
\begin{tabularx}{\textwidth}{lrrrrr}
\toprule
Rater & Valid N & Mean pref. & Tie rate & Majority agreement & Item-mean MAE \\
\midrule
Lowest-agreement rater (rater\_14) & 29 & 0.379 & 0.207 & 0.345 & 0.478 \\
Most extreme rater (rater\_11) & 30 & 0.450 & 0.100 & 0.433 & 0.465 \\
All raters mean & 29.9 & 0.441 & 0.343 & 0.481 & 0.311 \\
\bottomrule
\end{tabularx}

%% file: generated/natural_variance_summary.tex
% Natural-variance and null-result summary.
\begin{tabularx}{\textwidth}{p{0.22\textwidth} Y Y}
\toprule
Analysis & Statistic & Result \\
\midrule
Speaker gender & female vs. male speakers & no preference (n.s.; all $p \ge .21$) \\
Speaker accent & North-American vs. British speakers & no preference (n.s.) \\
Emotion gender interaction & gender $\times$ emotion & no interaction (n.s.) \\
Low-level descriptors & descriptor--score correlations & no corrected correlation (n.s.) \\
Model self-preference & Qwen judges on Qwen3-TTS vs. other TTS & no self-preference (n.s.) \\
Gemini TTS preference & TTS-system effect & preference differs by system (***) \\
Gemini-2.5 duration & duration--score correlation & negative correlation (***) \\
SQ-LLM rubric coupling & Emotional Impact vs. Overall Quality & low Emotional Impact is associated with more Overall=3 ratings (+16 pp; diagnostic) \\
Gemini rationale wording & expressive vs. monotone wording & score gap $\approx2.5$--$2.75$/10 (diagnostic) \\
\bottomrule
\end{tabularx}